\documentclass[a4paper,11pt]{article}
\usepackage{jcappub}
\usepackage[T1]{fontenc}
\usepackage{epsfig,fancyvrb,makeidx,latexsym,relsize}
\usepackage{amssymb}
\usepackage{mathrsfs}
\usepackage{slashed}
\usepackage{graphicx,color}
\usepackage{amsmath}
\usepackage{verbatim}

\def\lsim{\mathrel{\rlap{\lower4pt\hbox{\hskip1pt$\sim$}}
    \raise1pt\hbox{$<$}}}    
\def\gsim{\mathrel{\rlap{\lower4pt\hbox{\hskip1pt$\sim$}}
    \raise1pt\hbox{$>$}}}                
 
\title{Confronting the Galactic Center Gamma Ray Excess with a Light Scalar Dark Matter}

\author{Dilip Kumar Ghosh,}
\author{Subhadeep Mondal,}
\author{Ipsita Saha}
\affiliation{Department of Theoretical Physics, 
Indian Association for the Cultivation of Science, \\
2A \& 2B Raja S.C. Mullick Road, Kolkata 700032, India} 
\emailAdd{tpdkg@iacs.res.in}
\emailAdd{tpsm2@iacs.res.in}
\emailAdd{tpis@iacs.res.in}
\abstract{
The Fermi Large Area Telescope observed an excess in gamma ray emission spectrum coming from the center of the Milky Way galaxy.
This data reveals that a light Dark Matter (DM) candidate of mass in the range 31-40 GeV,
dominantly decaying into $b\bar b$ final state, can explain the presence of
the observed bump in photon energy. We try to interpret this observed phenomena by sneutrino DM annihilation
into pair of fermions in the Supersymmetric Inverse Seesaw Model (SISM). This model can also account for 
tiny non-zero neutrino masses satisfying existing neutrino oscillation data. We show that a Higgs portal DM in       
this model is in perfect agreement with this new interpretation besides satisfying all other existing collider, 
cosmological and low energy experimental constraints.
}

\begin{document}
\maketitle

\newpage

\section{Introduction:} \label{intro}
The existence of weakly-interacting massive Dark Matter (WIMP-DM) in the universe 
is by now a well established fact supported by various cosmological
and astrophysical observations. Several earth based underground experiments 
look for the direct evidence of WIMPs via its elastic scattering with 
nuclei of different target materials, the so called direct detection 
experiments. On the other hand, several indirect detection 
experiments have been going on to search for the products of WIMPs annihilation or 
decay which takes 
place either in the galactic halo/center, nearby galaxies or even in the 
Sun. Apart from these experiments, 
the Large Hadron Collider (LHC) plays a crucial role in scrutinizing
dark matter sector of several beyond the standard model (BSM) scenarios
by looking at the signal with large missing transverse energy.\\  

In this work, we are mainly interested in the indirect detection of the 
dark matter, in particular we are looking for the gamma ray signal arising 
from the annihilation of the DM in the center of the galaxy. The Fermi Large Area Telescope (LAT) 
is a satellite based experimental facility to study cosmic gamma-rays.
Over the last few years analyses with data obtained from Fermi gamma-ray telescope reported 
detection of a gamma-ray signal originated from central regions of the Milky way galaxy, with a 
high statistical significance \cite{Goodenough:2009gk,Hooper:2010mq,Hooper:2011ti,Abazajian:2012pn,
Gordon:2013vta,Hooper:2013rwa,Daylan:2014rsa,Abazajian:2014fta}. 
The energy spectrum and angular distribution of this signal 
indicates the existence of an extra component compatible with DM annihilation
into secondary photons coming from charged fermionic final states.
Analysis with the recent Fermi-LAT data suggests 
a best-fit DM candidate mass in the range $31 - 40~{\rm GeV}$ with $<\sigma v>_{b\bar b}$ 
with the annihilation cross-section $\simeq (1.4 - 2.0)\times 10^{-26}~{\rm cm^3.s^{-1}}$ \cite{Daylan:2014rsa}. 
The implications of these results have been further studied for various DM candidates 
in the context of several beyond the standard model (BSM) physics\cite{Okada:2013bna,Hagiwara:2013qya,
Anchordoqui:2013pta, Kyae:2013qna, Huang:2013apa,Modak:2013jya,
Boehm:2014hva, Alves:2014yha, Berlin:2014tja,Agrawal:2014una,Izaguirre:2014vva, Cerdeno:2014cda,
Ipek:2014gua, Kong:2014haa,Boehm:2014bia,Ko:2014gha,Abdullah:2014lla}.\\

It is an established fact that the standard model in its canonical 
form is unable to cater a dark matter candidate. Hence, 
to accommodate a DM particle one needs BSM physics. Several BSM
scenarios have been proposed with extra particle contents, which may 
provide a suitable candidate for the dark matter. The Minimal Supersymmetric Standard Model (MSSM) is 
one of the candidate scenarios for the BSM, which contains a very 
natural choice of a DM particle in the form of the Lightest Supersymmetric 
Particle (LSP). Within the paradigm of R-parity conserving MSSM \cite{Jungman:1995df,Griest:2000kj}, 
usually the lightest neutralino is the LSP and 
thereby the DM candidate. However, one should note that in constrained 
version of the MSSM (cMSSM), the lightest neutralino (LSP) lighter than $\sim 46~{\rm GeV}$ 
is excluded from the supersymmetric (SUSY) particle searches at the LEP \cite{pdg1}. 
On the other hand, the present SUSY particle search data from the LHC, pushes the lower 
bound on the LSP mass around $\gsim 200~{\rm GeV}$
\cite{pdg2,beskidt}. Obviously, the above mentioned galactic center gamma ray excess data 
can not be explained by this heavy neutralino dark matter particle, instead
one should look for another viable DM candidate, the sneutrino, 
the scalar superpartner of the left-handed neutrino in MSSM. The 
sneutrino being left handed has a very strong coupling with the $Z$-boson
which leads to very large annihilation and direct detection 
cross-section mainly via $Z$ mediated $s$ and $t$-channel diagrams 
respectively\footnote{Although the Higgs mediated s and t channel processes also play crucial
role here.}. This 
large cross-section immediately encounters severe constraints coming 
from the relic density and direct detection cross-section measurement
\cite{Falk:1994es}. Moreover a light left-handed sneutrino
$(m_{{\tilde \nu}_L} \leq \frac{M_Z}{2})$ is disallowed by the invisible decay
width of SM Z-boson \cite{Falk:1994es, Hebbeker:1999pi, Aprile:2012nq}. 
Nevertheless, this sneutrino dark matter scenario may still be viable, if one
introduces a right handed singlet chiral superfield, 
one for each family generation 
in the MSSM. This right handed neutrinos give rise to a Dirac mass 
to the SM neutrinos. Moreover, this theory being a supersymmetric, the 
right handed neutrino superfield contains its scalar partner which is 
a right handed sneutrino field. Now the physical sneutrino state is an
admixture of left(L) and right-handed(R) sneutrinos, (with reduced coupling
with $Z$ boson) which can now yield correct
relic density as well as direct detection cross-section. 
The primary motivation for introducing right handed neutrinos in the model 
is to generate tiny non-zero neutrino masses to satisfy the neutrino 
oscillation data which can not be explained within the framework of R-parity 
conserving MSSM. The simplest scenario is to introduce type-I seesaw mechanism 
\cite{seesawa, seesawb, seesawc, seesawd, seesawe} in MSSM. 
But this kind of models require either a very large Majorana 
mass term ${\cal O}(10^{15}~\rm GeV)$ or a very small Dirac Yukawa coupling ${\cal O}(10^{-5})$ to account for tiny neutrino masses,
which makes it phenomenologically less interesting. Here, we work with 
a different kind of seesaw scenario,
known as inverse seesaw mechanism \cite{inverse1, inverse2, Dev:2009aw, Dev:2010he,An:2011uq, 
Khalil:2011tb, Basso:2012ti}, which can bring down 
the seesaw scale considerably below TeV range with Dirac Yukawa couplings
as large as ${\cal O}(10^{-1})$. \\

In this paper, we concentrate on the supersymmetric version of the 
inverse seesaw scenario, the so called Supersymmetric Inverse Seesaw Model 
(SISM) \cite{Arina:2008bb, Hirsch:2009ra}, where one introduces two SM singlet
chiral superfields, one Dirac $N$ and one Majorana $S$ per family, 
to obtain the neutrino mass in the right ball park. 
Thus these models contain three lepton-number carrying electrically-neutral
fermions per family,
$(\nu_L, N^c, S)$. The physical neutrino state will be a linear superposition of these
three flavor states and their corresponding lightest scalar partners (sneutrino)could be a DM 
candidate with an admixture of both left and right handed 
sneutrinos. The possibility of a light scalar DM candidate under this 
scenario has already been studied in literature 
\cite{DeRomeri:2012qd, BhupalDev:2012ru, Banerjee:2013fga}. Whereas, one can indeed fit a light ($\sim 62~{\rm GeV}$) 
scalar DM in this scenario satisfying all the existing collider and DM constraints, it is also interesting to 
explore the possibility of having a lighter DM candidate that can explain the observed excess in gamma 
ray energy spectrum. In this work, we consider a DM mass in $31-40~{\rm GeV}$ region to confront the galactic center 
gamma-ray excess. Here, a pair of DM 
sneutrinos annihilate into a pair of bottom quarks or tau leptons mediated by 
Higgs bosons to produce correct relic density, with an annihilation  cross-section 
$\sim 5.0\times10^{-27}~\rm{cm^3.s^{-1}}$.
It is worth mentioning that this dark matter mass window is still allowed by the latest direct detection
limits from the LUX experiment~\cite{Akerib:2013tjd}. For 40 GeV sneutrino DM scenario our 
model predicts the existence of a lighter Higgs boson of mass 86.7 GeV with heavier Higgs boson at 125 GeV
satisfying all the experimental data.\\

The paper is organized as follows. In Section~\ref{model} we discuss
the sneutrino dark matter scenario in SISM and set up our notations. 
In Section~\ref{fermi} we discuss our strategy to fit the gamma ray excess data 
in SISM. In this 
section we also comment on different low energy constraints used in our 
analysis. In Section~\ref{chisq_anal} we do the 
$\chi^2$ analysis of the model parameters in the light of the 
galactic center gamma ray spectrum and also 
present the main results of our analysis. Finally 
we conclude in Section~\ref{concl}. 
\section{$\tilde \nu$ LSP scenario in SISM:} \label{model}
It is an well established fact that neutrinos have tiny mass and in order
to generate such a tiny mass for them one needs to invoke physics beyond the
standard model. The supersymmetric inverse seesaw mechanism (SISM) 
is one such possibilities, where one adds three 
SM singlet (Dirac) $\hat{N}^c_i$ and (Majorana) $\hat{S}_i$ (with $i=1,2,3$) 
superfields with lepton number $-1$ and $+1$ respectively to the MSSM 
field contents. 
The superpotential is given by
\begin{eqnarray}
	{\cal W}_{\rm SISM}={\cal W}_{\rm MSSM}+\epsilon_{ab}y_\nu^{ij}\hat{L}^a_i\hat{H}^b_u \hat{N}^c_j+ M_{R_{ij}}
\hat{N}^c_i\hat{S}_j+\mu_{S_{ij}}\hat{S}_i\hat{S}_j \, ,
\label{sup}
\end{eqnarray}
where, $\mu_S$ is a small lepton number violating parameter. The soft 
SUSY-breaking Lagrangian is  
\begin{eqnarray}
	{\cal L}_{\rm SISM}^{\rm soft} &=& {\cal L}_{\rm MSSM}^{\rm soft} -\left[m_N^2\widetilde{N}^{c^{\dag}}\widetilde{N}^c+
	m_S^2\widetilde{S}^\dag\widetilde{S}\right]\nonumber\\
	&& -\left[\epsilon_{ab}A_\nu^{ij}\widetilde{L}^a_i\widetilde{N}^c_jH_u^b+B^{ij}_{M_R}\widetilde{N}^c_i\widetilde{S}_j+
	B_{\mu_S}^{ij}\widetilde{S}_i\widetilde{S}_j+{\rm h.c.}\right]. 
\label{soft}
\end{eqnarray}
The $9\times 9$ tree level neutrino mass matrix in the basis $\{\nu_L, N^c, S\}$ is: 
\begin{eqnarray}
	{\cal M}_\nu = \left(\begin{array}{ccc}
		{\bf 0} & M_D & {\bf 0}\\
		M_D^T & {\bf 0} & M_R \\
		{\bf 0} & M_R^T & \mu_S
	\end{array}\right),
	\label{eq:mbig}
\end{eqnarray}
where $M_D = v_u y_{\nu}$ is the Dirac neutrino mass matrix, $v_u=v\sin\beta$ being the vacuum expectation value (vev) 
of the $\hat H_u$ superfield in MSSM, with $v\simeq 174$ GeV. Under the approximation,  
$\|\mu_S\|\ll \|M_R\|$ (where $\|M\|\equiv \sqrt{{\rm Tr}(M^\dag M)}$), we can extract the $3\times 3$ 
light neutrino mass matrix: 
\begin{eqnarray}
	M_\nu = \left[M_DM_R^{T^{-1}}\right]\mu_S\left[(M_R^{-1})M_D^T\right]+{\cal O}(\mu_S^2) \equiv F\mu_S F^T+{\cal O}(\mu_S^2) \, .
	\label{eq:vmass}
\end{eqnarray}
As evident from Eq.~(\ref{eq:vmass}), the smallness of neutrino mass now depends on the small lepton-number violating parameter 
$\mu_S$ instead of just the smallness of the Dirac mass $M_D$ and/or heaviness of $M_R$ as 
in the canonical type-I seesaw case. As a result, we can easily have a $M_R$ 
below the ${\rm TeV}$ range even with a comparatively large Dirac Yukawa coupling ($\sim 0.1$). This feature makes the model 
particularly interesting phenomenologically~\cite{BhupalDev:2012ru, Chen:2011hc, Mondal:2012jv, Das:2012ze, 
Bandyopadhyay:2012px,Banerjee:2013fga,Guo:2013sna,Arina:2013zca}.  

We fit the neutrino oscillation data in the model by an off-diagonal $\mu_s$, while keeping $m_D$ and $M_R$ strictly 
diagonal. We also consider the corresponding
$A_{\nu}$ and $B_{M_R}$ to be diagonal. This diagonal structure of these matrices 
reduces the number of free parameters as well as keeps the Lepton Flavor Violating (LFV) processes well within the experimental limits. \\
Due to mixing between doublet and singlet sneutrinos we have a $9\times 9$ complex sneutrino mass squared matrix in the theory. 
We assume CP conservation in the soft SUSY breaking Lagrangian. As a result, we can easily break this mass matrix into block-diagonal
form with two $9\times 9$ real matrices corresponding to CP-even and CP-odd sneutrino states. 
The corresponding mass term looks like
\begin{eqnarray}
 {\cal L}_{\tilde\nu} = \frac{1}{2} (\phi^R, \phi^I) \left(\begin{array}{cc}
		{\cal M}_+^2 & {\bf 0} \\
		{\bf 0} & {\cal M}_-^2
	\end{array}\right) \left(\begin{array}{c}{\phi^R}\\ {\phi^I}\end{array}\right),
\end{eqnarray}
where $\phi^{R,I} = (\widetilde\nu^{R,I}_{L_i},\widetilde N^{c^{R,I}}_j,\widetilde S^{R,I}_k)~ 
(i,j,k = 1, 2, 3)$ and  
\begin{eqnarray}
{\cal M}_\pm^2 = \left(\begin{array}{ccc}
		m_{\tilde L}^2+M_DM_D^T+\frac{1}{2}m_Z^2\cos 2\beta & \pm(v_uA_\nu-\mu M_D\cot\beta) & M_DM_R\\
		\pm(v_uA_\nu-\mu M_D\cot\beta)^T & m_N^2+M_RM_R^T+M_D^TM_D & B_{M_R}\pm M_R\mu_S\\
		M_R^TM_D^T & B^T_{M_R}\pm \mu_SM_R^T & m_S^2+\mu_S^2+M_R^TM_R\pm B_{\mu_S}
	\end{array}\right), \nonumber
	\label{eq:svmass}
\end{eqnarray}
where, $m_{\tilde L}^2$ is the soft SUSY-breaking mass squared term for $SU(2)_L$-doublet sleptons. One can diagonalize 
the real symmetric CP-even and CP-odd mass squared matrices ${\cal M}_\pm^2$ by $9\times 9$ orthogonal 
matrices ${\cal G}_\pm$:
\begin{eqnarray} 
 {\cal G}_\pm {\cal M}^2_\pm {\cal G}^T_\pm = {\rm diag} \left(m^2_{\widetilde\nu^{R,I}_i}\right)~~~~(i=1, 2,\cdots, 9).
\end{eqnarray} 
In the present work, we choose the parameter space such that the lightest mass eigenstate of the mixed sneutrinos is 
the LSP in the model and thereby the candidate for a stable Dark Matter. The degeneracy between the eigenvalues 
corresponding to ${\cal M}^2_\pm$ are lifted due to the lepton number breaking parameter, $\mu_s$.  After fitting 
neutrino oscillation data in the model and choosing the parameter space accordingly to fit a suitable scalar DM 
candidate, we obtain a $\mu_s$ at most in keV order. This creates a small non-degeneracy in sneutrino masses obtained 
from two block-diagonal matrices in keV order. These eigenstates are, therefore, considered degenerate for the present study.   
  
\section{Confronting the Galactic Center Gamma-ray Excess:} \label{fermi}
The Fermi-LAT experiment is devoted towards the study of the entire gamma-ray sky consisting of photons with energy ranging
from few MeVs to several hundred GeV. Several independent studies of the obtained data obtained an excess of gamma-rays 
peaking at $\sim 2-3~{\rm GeV}$, originating from the galactic center. The excess cannot be explained by any known 
astrophysical sources so far. 
It has been argued that the signal can be well interpreted 
with a DM of mass in between 31-40 GeV \cite{Daylan:2014rsa}
annihilating to $b\bar b$ final state with an annihilation cross-section $<\sigma v> = (1.4~-~2.0)\times 10^{-26}~{\rm cm^3s^{-1}}$. \\

Here we try to explain the observed gamma ray excess in our model with a 31-40 GeV sneutrino DM. 
We do a $\chi^2$ analysis to find the best-fit model parameter values and corresponding 
annihilation cross-section that fit the photon spectrum. Before we perform the detailed $\chi^2$ analysis
we first give a detailed account of different ranges of SISM parameters used in our analysis.
It is important to have a right mixing of $\rm L$ and $\rm R$- handed components in the LSP state to 
satisfy different DM constraints. A sizable left-handed component in the sneutrino
can give rise to a large direct detection cross-section due to an enhanced coupling with Z-boson,
which is in direct conflict with the latest experimental data. The L-R mixing 
is controlled by the parameters $y_{\nu}$ and $A_{\nu}$ appearing in the off-diagonal entries of the sneutrino mixing matrix.
Hence we are forced to keep these mixing parameters small. The $A_{\nu}$ parameter also 
controls the trilinear Higgs-sneutrino-sneutrino coupling and cannot be large so as to prevent too much invisible Higgs decay 
width which is tightly constrained \cite{Banerjee:2013fga} from the most recent Higgs boson data published by the ATLAS and CMS
collaborations.
We choose all the A-terms in this model to be negative, however, it is to be noted that a positive $A_{\nu}$
is also suitable for this study. In our choice of parameter space, 
for $m_h \gsim 2 m_{\tilde \nu}$,  the Higgs invisible branching ratio $(h \rightarrow \tilde \nu \tilde \nu)$ is always less than $1\%$ 
due to suppressed trilinear scalar coupling.
The choice of $B_{M_R}$ also affects the composition of the LSP sneutrino states significantly. 
We observe that for $(M_R)_{11}$ in the range 100-1000 GeV and the elements of $m_S$ and $m_N$ fixed at 1000 GeV, 
to get a sufficiently light DM mass, 
we need to keep $(B_{M_R})_{11}$ in the (1.0 - 2.0)$\times 10^{6}~{\rm GeV^2}$ region.   
Smaller choice of $(B_{M_R})_{11}$ requires a smaller value of $(M_R)_{11}$ to have 
a sneutrino LSP in our desired mass range. Note that, by lowering the value of 
$(M_R)_{11}$ one can hit a situation where the heavy neutrino masses
become lighter than $\frac{M_Z}{2}$, which is ruled out by the LEP measurement
of the invisible decay width of the $Z$ boson. On the other hand, if we choose a 
larger $(B_{M_R})_{11}$, we get dark matter mass in the right window of our choice provided
$(M_R)_{11}$ is also very large. But this large $(M_R)_{11}$ will in turn decrease the contribution 
of the right-handed component and increase that of the left-handed ones 
which is disfavored from 
the DM direct detection limits as mentioned previously. However, this large 
$(M_R)_{11}$ scenario can be saved if we consider a smaller $(y_{\nu})_{11}$ and $(A_{\nu})_{11}$ 
which will reduce the annihilation cross-section $<\sigma v>_{b \bar b}$ considerably due 
to suppressed sneutrino couplings to Higgs bosons. We don't take a large tan$\beta$ in order to keep anti-proton flux 
coming from DM annihilation small.  

As explained in Section~\ref{model}, the  
$M_R$, $B_{M_R}$, $y_{\nu}$ and $A_{\nu}$, are all diagonal, as a result of this, the sneutrino (DM) mass and its couplings 
strongly depends on the (1,1) element of these matrices. 
Below we show the parameter ranges of our scan  
to obtain a suitable sneutrino DM candidate with a correct mass
and composition:
\begin{eqnarray}
 &&\tan \beta \in [5,20] ~;~ (M_R)_{11} \in [100,1000]~\rm GeV~; \nonumber \\
 && (y_{\nu})_{11} \in [0.01,0.10]~;~  (B_{M_R})_{11} \in [1.0,2.0]\times 10^6 ~\rm (GeV)^2~ ; \nonumber \\
 && (A_{\nu})_{11} \in [-1000,-1] ~\rm GeV~; 
\label{param-range}
 \end{eqnarray}
 To perform the SISM parameter space scanning we first implement the SISM Lagrangian in the SARAH-3.3.2\cite{sarah} code to obtain the 
 appropriate model file for the SPheno-3.2.1\cite{spheno} and micrOMEGAs-2.4.5 \cite{micromegas}. We then use 
 SPheno-3.2.1\cite{spheno} to generate the spectrum and different low energy constraints on SISM parameter space. 
 Finally, we run the micrOMEGAs-2.4.5 to calculate different dark matter observables, e.g.
 direct detection cross-section, annihilation rate etc. relevant for our analysis. 
  Throughout our study we fix the Navarro-Frenk-White (NFW) density profile \cite{Navarro:1995iw,Navarro:1996gj}
  for the DM density distribution.
  Below we provide detailed list of constraints imposed while scanning the SISM parameter space : 
\begin{itemize}
 
\item Mass of one of the two CP-even neutral Higgs bosons is always set within the $3\sigma$ range of the ATLAS-CMS combined
best fit value, $125.6~{\rm GeV}$~\cite{cms:higgs,atlas:higgs}. Squark, gluino and gaugino mass limits are also taken into account 
according to the latest LHC results~\cite{atlas:sparticle,cms:sparticle}. 
 
 \item ${\rm BR}(B_s\rightarrow \mu^+ \mu^-)$ is a flavor physics constraint that puts strong bound on the MSSM parameter space. It
 is inversely proportional to $m_A^4$ and proportional to $(\tan\beta)^6$. Thus this constraint may be severe 
 for the case with lighter Higgs boson states and large $\tan\beta$. We chose to work with a small $\tan\beta$ and $m_A \gsim 150$ GeV 
for all our benchmark points as we explain in the later section.
 We stay within the $2\sigma$ reach of the most recent LHCb measurement of 
 ${\rm BR}(B_s\rightarrow \mu^+ \mu^-)= (2.9^{+1.1}_{-1.0})\times 10^{-9}$ \cite{Aaij:2013aka, Chatrchyan:2013bka}.
 
 \item The fact that measurement of BR($b\rightarrow s\gamma$) agrees quite well with the predicted value from SM, makes it very effective 
 to constrain any BSM parameter space. Within the framework of MSSM, dominant contribution comes from charged Higgs and chargino exchange 
 diagrams. These contributions interfere destructively only if $\mu$ and $A_t$ are of opposite sign. We chose to work with a positive $\mu$
 and negative $A_t$. We consider the $2\sigma$ range of ${\rm BR}(b\rightarrow s\gamma) = (3.21\pm 0.33)\times 10^{-4}$ 
 \cite{Lees:2012ym} for this purpose.
 
 \item Stringent constraints can come from the lepton flavor violating (LFV) decay branching ratios \cite{pdg2012} in case of a large 
 off-diagonal Dirac  neutrino Yukawa coupling ($y_{\nu}$). For the present purpose, we chose to work with a strictly diagonal $y_{\nu}$. 
 Hence, there is no excess in the LFV decay branching fractions. 
 
 \item The SUSY contribution to the lepton anomalous magnetic moment is defined by the difference between the experimental result and the SM
prediction of the magnetic moment of lepton ($\delta a_{\ell} $ = $a^{\rm expt}_{\ell}$ - $a^{\rm SM}_{\ell}$). These contributions 
may be enhanced due to the 
chargino-sneutrino-charged lepton or neutralino-slepton-charged lepton loop contributions. The Muon anomalous magnetic moment 
 \cite{Venanzoni:2012yp} is more important one, showing a $3\sigma$ discrepancy \cite{Hagiwara:2003da} over the SM value which has large hadronic uncertainty. 
For the electron anomalous magnetic moment \cite{Aoyama:2012wj}, the discrepancy is quite small. 
We do not consider tau anomalous magnetic moment.
 
 \item Because of the mixing among the light and heavy neutrino states, non-unitarity can be a crucial constraint 
 \cite{Abada:2007ux, Antusch:2008tz} in neutrino mass 
 models like these. Non-unitarity parameters ($|\eta|_{\ell_i \ell_j}$, where $\ell_{i,j}$ denotes $e$, $\mu$ or $\tau$) are derived from  
 neutrino oscillation data, the LEP precision data, weak gauge boson decays and the lepton-flavor violating (LFV) decays.
 Since we work with strictly diagonal $M_R$ and $y_{\nu}$ matrices, the non-unitarity parameters will also construct a diagonal matrix. 
 
 \item For relic density, we take the best fit value provided by the PLANCK experiment \cite{Ade:2013zuv}, $0.1199\pm 0.0027$. 
 We fit the relic density 
 value within the $2\sigma$ error bar of the best fit value. In the present scenario, we obtain the proper relic density only because of 
 the s-channel resonance effect. The dominant annihilation channel is $b\bar b$. The second most contributing channel to 
 the relic density is $\tau\bar\tau$.  
 
 \item XENON100 \cite{Aprile:2012nq} and subsequently LUX \cite{Akerib:2013tjd} experiment have put upper-bound on the DM-nucleon elastic scattering cross-section 
  for a wide range of DM mass. It is very difficult to stay 
 below the upper limit with a sizable left-handed component in the sneutrino LSP state. Hence our DM candidate is dominantly right-handed.
 Future experiments like XENON1T \cite{Aprile:2012zx} can probe a direct detection cross-section at most in $10^{-11}~{\rm pb}$ range. 
 
 \item A large DM pair annihilation cross-section into $b\bar b$ channel means the annihilation cross-section into other quark pairs should 
 also increase. As a result, the anti-proton flux in the final state may increase. We also take into account the most recent PAMELA data \cite{Adriani:2008zq} for anti-proton flux to compare with the same in our model. 
 
 \end{itemize}
We are now in a position to discuss the feasibility of a light sneutrino DM candidate in SISM in 31-40 GeV mass range decaying 
into fermionic final states to account for the galactic center gamma ray excess.  
\subsection{Relic Density \& Annihilation Cross-section:}
In this section we discuss about the relic density and the annihilation 
cross-section for the dark matter mass range, 31-40 GeV. \\
 
\begin{figure}[ht]
\centering
\includegraphics[height=6cm,width=7cm]{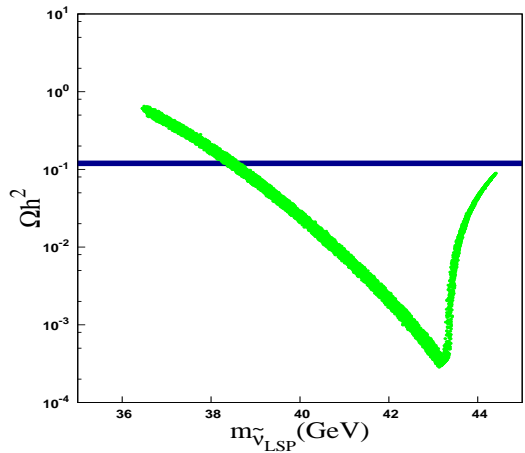}
\caption{Relic density distribution (green band/light gray) 
as a function of the DM mass. The horizontal blue (dark) band 
indicates the $2\sigma$ allowed region around the best fit value given by the PLANCK.}
\label{fig:light_DM_relic_density}
\end{figure}

\begin{figure}[ht]
\centering
\includegraphics[height=6cm,width=7cm]{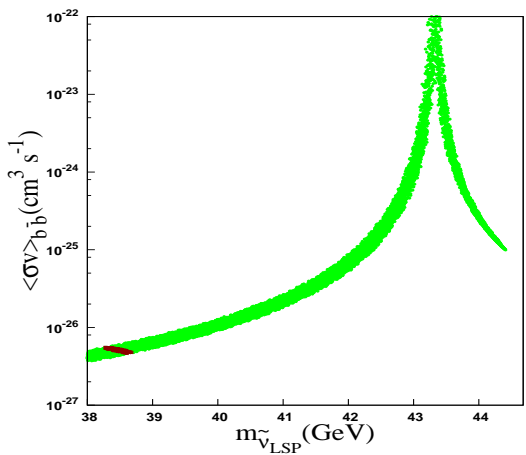}
\caption{The variation of the annihilation cross-section in 
$b\bar b$ channel with the DM mass around the resonance is shown by
the green (light gray) band. The red (dark) patche is the only 
region where the relic density lies within $2\sigma$ 
of the best-fit value.}
\label{fig:light_DM_sigv_bb}
\end{figure}

As evident from Eq.~(\ref{soft}), the
sneutrino states couple with the Higgs bosons via the trilinear scalar coupling $A_{\nu}$. 
On the other hand, $b\bar b$ is the most preferred decay channel for the Higgs bosons. As a result of these the 
DM pair annihilation into $b\bar b$ final state through Higgs boson mediated s-channel process is the most dominant one.
Note that, the sneutrino LSP being dominantly right handed, its coupling with the Z-boson is negligibly small which 
makes the Z-mediated annihilation channel contribution redundant. 
Even in the absence of a large $A_{\nu}$, one can enhance the annihilation cross-section in this decay channel 
by making $m_{\tilde\nu}\simeq \frac{m_h}{2}$, where, $m_h$ the mass of the Higgs boson, thereby
producing a resonating effect in the annihilation channel which is essential to produce correct relic abundance 
with a sneutrino DM. Now for lightest Higgs boson mass with $m_h =125 $ GeV and other Higgs bosons decoupled from the
spectrum, it would not be possible to have $31-40~{\rm GeV}$ sneutrino DM as it will 
result in a relic overabundance due to non-availability of resonance channel for sneutrino annihilation process\footnote{Note that all other 
superparticles are very heavy $\gsim {\cal O}$(TeV) thus 
their effects in DM interactions can be neglected as far as this analysis 
is concerned.}.
Therefore, with 31-40 GeV sneutrino DM, one requires at least one Higgs boson with mass in the vicinity of 
$60-80~{\rm GeV}$ \cite{Berlin:2014tja} to produce the s-channel resonance in the annihilation channel to yield proper relic density.
We scan our MSSM like Higgs sector using HiggsBound-4.1.0\cite{Bechtle:2008jh} package and obtain
light CP even Higgs boson with mass as light as $86.7~{\rm GeV}$ allowed by LEP, Tevatron and Higgs boson 
search data both at 7 and 8 TeV LHC run. We should mention here that the heavier CP even 
Higgs boson lies within $2\sigma $ range of the best-fit value of the Higgs mass measured at the LHC and 
the corresponding charged Higgs and CP-odd Higgs bosons are heavier than 150 GeV. One should note that a CP-even 
Higgs boson with a mass lighter than 86.7 GeV is excluded from the Higgs search at LEP in the channel~\cite{Barate:2003sz}
\begin{eqnarray}
 e^+ e^- \rightarrow Z h \rightarrow Z b \bar b, 
\end{eqnarray}
as obtained from the HiggsBound code.
From the above discussion, it is evident that in order to raise the value of $<\sigma v>_{b\bar b}$ to 
$10^{-26}~{\rm cm^3.s^{-1}}$ one requires to have a DM candidate with a mass very 
close to $m_h/2$. In this scenario because of low lying Higgs mass spectrum
($m_h = 86.7~{\rm GeV},m_H = 125~{\rm GeV}, m_A \approx m_{H^\pm} \geq 150
~{\rm GeV}$), all these Higgs states may contribute 
to the annihilation process. However, the resonance dip in Fig.~\ref{fig:light_DM_relic_density} suggests that the 
most dominating contribution comes from the lightest Higgs boson mediated channel. 
In this figure, we show the distribution of relic density as a function of the DM mass. The thickness of the relic density 
band correspond to the random variation of different model parameters that we scan. The horizontal blue (dark) band  
indicates the $2\sigma$ allowed region around the best fit value given by the PLANCK \cite{Ade:2013zuv}. As can be seen,  
the relic density constraint is satisfied for a DM mass which is little away from $m_h/2$, where 
$m_h = 86.7$ GeV is the lightest Higgs boson mass. At the resonance point the annihilation rate is too large which 
results into under abundance of the sneutrino DM. In Table \ref{tab:light_higgs_relic_contbn}, we show the relative contributions 
of the dominant channels to get proper relic density for this DM mass range. 
It is clear that the sneutrino pairs dominantly annihilate into 
a $b\bar b$ final state and to the $\tau\bar\tau$ final state with $86\%$ and 
$14\%$ probabilities respectively.
\begin{table}[h!]
\begin{center}
\begin{tabular}{||c|c||}\hline\hline
Dominant channels & Relative contribution   \\ 
\hline\hline
$\widetilde\nu_1 \widetilde\nu_1 \to b\bar b$ & $86\%$  \\ 
$\widetilde\nu_1 \widetilde\nu_1 \to \tau\bar\tau$ & $14\%$  \\
\hline\hline 
\end{tabular}
\end{center}
\caption{The relative contributions to $\frac{1}{\Omega h^2}$ coming from 
dominant annihilation channels for a 
sneutrino DM mass close to $40~{\rm GeV}$.}
\label{tab:light_higgs_relic_contbn}
\end{table}
In Fig.~\ref{fig:light_DM_sigv_bb} we show $<\sigma v>_{b\bar b}$ distribution as a function of the DM mass. 
The red patch corresponds to the DM mass range that satisfy the relic density 
constraint ($0.1145<\Omega h^2<0.1253$) as shown in Fig.~\ref{fig:light_DM_relic_density}.
As expected, the annihilation cross-section is much higher at the resonance and the allowed parameter space is tightly 
constrained by relic density. In this case the annihilation 
cross section is s-wave dominated and it can be expressed as 
\cite{Wells:1994qy}: 
\begin{eqnarray}
\sigma v_{{\rm rel}}=a_{{\rm rel}}+b_{{\rm rel}}v^2_{{\rm rel}},
\end{eqnarray}
with,
\begin{eqnarray}
&&a_{{\rm rel}}\simeq\frac{3A^2m_b^2sin^2\alpha(m_{\widetilde\nu_1}^2-m_b^2)^{3/2}}
{4\pi m^3_{\widetilde\nu_1}v^2_ucos^2\beta(4m^2_{\widetilde\nu_1}-m_h^2)^2} \\
&&b_{{\rm rel}}\simeq \frac{3A^2m_b^2sin^2\alpha(2m_b^2-m_{\widetilde\nu_1}^2)\sqrt{m^2_{\widetilde\nu_1}-m_b^2}}
{32\pi m^3_{\widetilde\nu_1}v_u^2cos^2\beta(4m^2_{\widetilde\nu_1}-m^2_h)^2} - 
\frac{3A^2m_b^2sin^2\alpha(m^2_{\widetilde\nu_1}-m^2_b)^{3/2}}
{128\pi m^3_{\widetilde\nu_1}v^2_ucos^2\beta(4m^2_{\widetilde\nu_1}-m^2_h)^2} +   \nonumber \\
&&\frac{3A^2m_b^2sin^2\alpha\sqrt{m_{\widetilde\nu_1}^2-m_b^2}}
{16\pi m_{\widetilde\nu_1}v_u^2cos^2\beta(4m^2_{\widetilde\nu_1}-m^2_h)^2}
\Big(1-8\frac{m_{\widetilde\nu_1}^2-m_b^2}{4m_{\widetilde\nu_1}^2-m_h^2}\Big)
\end{eqnarray}
where, $A$ denotes the Higgs-DM-DM coupling, $m_b$ denotes the bottom quark mass and $v_u$ denotes the vacuum expectation value 
of up-type Higgs boson. We compute the annihilation cross-section using micrOMEGAs.

\begin{figure}[ht]
\centering
\includegraphics[height=6cm,width=7cm]{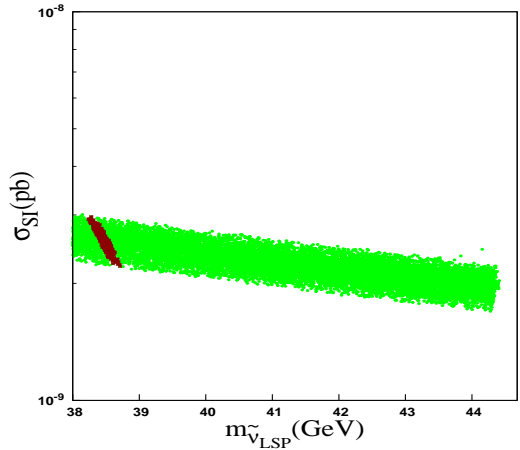}
\caption{The variation of the spin-independent cross-section with the DM mass around the resonance is shown. 
The red points also satisfy the relic density constraint.}
\label{fig:light_DM_sig_SI}
\end{figure}

The value of spin-independent cross-section, $\sigma_{SI}$, for light DM candidates is highly constrained    
from the latest LUX results \cite{Akerib:2013tjd}. In this model, the sneutrino DM being singlet-like, couples very feebly with the Z-boson 
via the extremely small left-handed component. The dominant contribution to direct detection cross-section, therefore, again comes from the 
light-Higgs mediated channel. The DM-nucleon elastic scattering cross-section is given by
\begin{eqnarray}
\sigma_{SI}^n \simeq \frac{A^2}{\pi v_u^2cos^2\beta}f^2\frac{m_p^4}{m_h^4(m_{\widetilde\nu_1}+m_p)^2},
\end{eqnarray}
where, $m_p$ denotes the proton mass and $f$ appears in the Higgs-nucleon-nucleon coupling and is $\sim 0.3$. 
We observe that $\sigma_{SI}$ for our relevant parameter space is slightly above the best-fit limit provided by the LUX experiment 
as evident in Fig~\ref{fig:light_DM_sig_SI}.
However, one should note that there exist various uncertainties mainly arising from 
from particle physics and astrophysics related issues 
\cite{Gondolo:2013xya,Ellis:2008hf} in the computation of $\sigma_{SI}$. 
Uncertainties arise from the determination of local DM density \cite{Beskidt:2012bh,Bovy:2012tw}. 
Also assuming non-Maxwellian velocity distributions for WIMP gives rise to significant variation in the direct detection 
rate \cite{Fairbairn:2012zs,Bhattacharjee:2012xm} specially for low DM masses. Considering all these uncertainties, one can relax 
the upper bound on the existing direct detection cross-section upto one order 
of magnitude, thus allowing our model prediction of $\sigma_{SI}$. We expect that in future run of the LUX 
experiment should be able to provide a good test of this parameter space.  
   
Hence, our analysis so far indicates that in this model one can have a $\sim 40~{\rm GeV}$ sneutrino DM candidate with
significant annihilation cross-section into $b\bar b$ final state satisfying relic density and 
direct detection cross-section constraints.  
Now it is left to be seen if we can explain the galactic center gamma ray excess with this DM.  
In the next section, we carry out a $\chi^2$ minimization considering the photon fluxes as experimental data points to find the best
fit model parameters to account for the excess in the spectrum. 
\section{$\chi^2$ analysis and Results:}\label{chisq_anal}
\begin{figure}
\centering
\includegraphics[trim=0 100 10 100,clip,height=10cm,width=10cm]{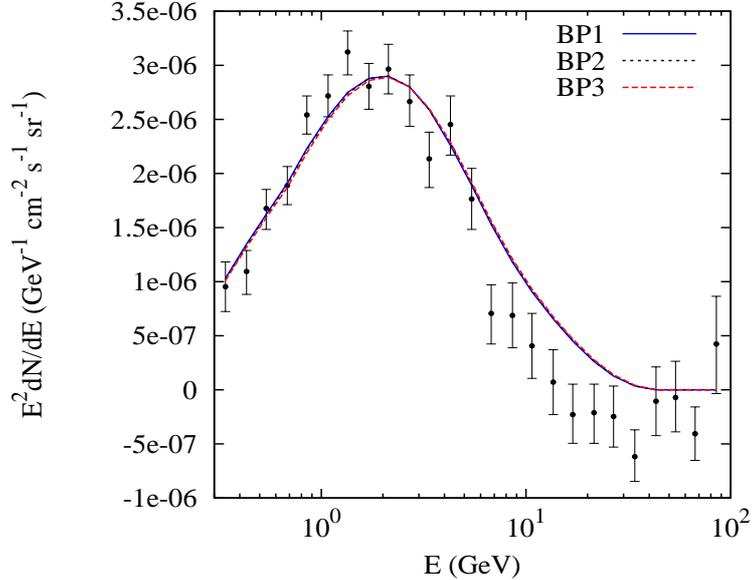}
\caption{The best fit signal photon spectrum in the latitude $|b|<5^{\circ}$ for galactic center gamma-ray excess. 
The black points represent the gamma ray excess data of the photon flux. The 
solid magenta, dashed black, dotted red and dash-dotted blue lines show the 
fit for the SISM benchmark points BP1, BP2 and BP3 respectively.}
\label{spec_fit}
\end{figure}

In this section, we compare our results for the gamma ray spectrum 
corresponding to the galactic center excess in the $|b|<5^{\circ}$ galactic latitude slice, with the
data and corresponding uncertainties are taken from \cite{Daylan:2014rsa}.
The $\chi^2$ is defined as, 
\begin{eqnarray}
 \chi^2 = \sum_i \frac{(\Phi_i^{model} - \Phi_i^{observed})^2}{\delta\Phi_i^2} \nonumber
\end{eqnarray}
where, $i$ runs for all the data points, $\Phi_i$'s are the photon fluxes i.e.
the observables in this case, $\delta{\Phi_i}$ are the 
corresponding experimental errors.
We use only the non-negative data points (16 in total) in the $\chi^2$ analysis.
As mentioned before, $(M_R)_{11}, (B_{M_R})_{11}, (y_{\nu})_{11}$ and $(A_{\nu})_{11}$
are the most relevant parameters that control the DM mass and its mixing. 
for this analysis, we fix the $(B_{M_R})_{11}$ parameter in the 
range mentioned in Eq.~\ref{param-range}. We then vary the 
three model parameters, namely, $(M_R)_{11}, (y_{\nu})_{11}$ and $(A_{\nu})_{11}$ 
to generate the signal photon spectrum, which we fit against the galactic center gamma ray excess data. \\ 
For our illustration purpose, we provide three benchmark points with different $({B_{M_R}})_{11}$ values.
In Table \ref{chisq}, we show the best-fit three model parameters and the corresponding $\chi^2$ 
per degrees of freedom (d.o.f = 13). It is to be noted that the 
goodness of the fit remains almost same even for the three distinct model parameter sets. \\
\begin{table}[htb!]
\begin{center}
\begin{tabular}{|c|c|c|c|c|}\hline\hline 
& BP1 & BP2 & BP3 \\ 
\hline\hline
$(M_R)_{11}~({\rm GeV})$ & 787.99 & 634.42 & 448.69 \\
$(y_{\nu})_{11}$ & 0.0467 & 0.0328 & 0.0725 \\
$(A_{\nu})_{11}~({\rm GeV})$ & -174.35 & -140.11 & -84.30 \\
$\chi^2_{min}$ & 14.069 & 14.441 & 15.481 \\
$\chi^2_{min}/d.o.f$ & 1.08 & 1.11 & 1.19 \\
\hline\hline 
\end{tabular}
\end{center}
\caption{Best-fit parameter values at minimum $\chi^2/d.o.f$.}
\label{chisq}
\end{table}
In Fig.~\ref{spec_fit}, we show our best-fit signal gamma-ray spectrum in the latitude 
$|b|<5^{\circ}$ for our three benchmark points. It is worth noting that these fits correspond to
$<\sigma v>_{b\bar b} \sim 5\times 10^{-27}~{\rm cm^3s^{-1}}$, which is roughly half of the limit given 
in \cite{Daylan:2014rsa}. \\ 
Now we discuss the results of our analysis in different part of the parameter space and see if we can put some 
constraints on the model parameters. 
Considering all the before mentioned constraints, we take the inputs given in Table \ref{tab:low} for our benchmark points. 
Note that apart from tan$\beta$, the only change in the input parameters in different benchmark points are the (1,1) elements 
of all the matrices $y_{\nu}$, $M_R$, $A_{\nu}$ and $B_{M_R}$. As a result of the assumed diagonal texture for all these matrices,
DM mass and mixing are sensitive only to these elements. 
The corresponding values of the constraints are also provided in Table \ref{tab:low}. 
\begin{table}[h!]
\scriptsize
\begin{tabular}{||c|c|c|c|c||} \hline\hline
Parameters/Observables & BP1 & BP2 & BP3 \\ \hline\hline
$\tan\beta$ & 10.0 & 11.7 & 11.5 \\ 
$y_\nu$ & (0.0467,0.01,0.01) & (0.0328,0.01,0.01) & (0.0725,0.010,0.010) \\
$M_R$ (GeV) & (787.99,1000,1000) & (634.42,1000,1000) & (448.69,1000,1000) \\
$A_{\nu}$ (GeV) & -(174.35,1.0,1.0) & -(140.11,1.0,1.0) & -(84.30,1.0,1.0) \\
$B_{M_R}$ (GeV$^2$) & (1.53,1.0,1.0)$\times 10^6$ & (1.40,1.0,1.0)$\times 10^6$ & (1.20,1.0,1.0)$\times 10^6$ \\ \hline
$m_h$ (GeV) &  86.7 & 86.8 & 88.0 \\ 
$m_H$ (GeV) & 126.5 & 126.2 & 126.3  \\ 
$m_A$ (GeV) & 203.8 & 224.2 & 223.5  \\ 
$m_{\widetilde\nu_1}$ (GeV) & 38.4 & 38.8 & 39.4  \\\hline
$\Omega_{\rm DM}h^2$ & 0.127 & 0.125 &  0.127 \\ 
$\sigma_{\rm SI}$ (pb) & $2.52\times 10^{-09}$ & $2.09\times 10^{-09}$ & $2.05\times 10^{-09}$ \\ 
\hline
$\delta a_\mu$ & $2.3\times 10^{-10}$ & $6.0\times 10^{-10}$ & $5.9\times 10^{-10}$ \\
$\delta a_e$ & $1.2\times 10^{-15}$ &  $1.3\times 10^{-15}$ & $1.2\times 10^{-15}$ \\ \hline
BR$(B\to X_s\gamma)$ & $3.1\times 10^{-4}$ & $3.0\times 10^{-4}$ & $3.0\times 10^{-4}$ \\
BR$(B_s\to \mu^+\mu^-)$ & $4.3\times 10^{-9}$ & $4.7\times 10^{-9}$ & $4.6\times 10^{-9}$ \\ \hline
BR$(\mu\to e\gamma)$ & $1.1\times 10^{-23}$ & $1.3\times 10^{-23}$ & $1.3\times 10^{-23}$ \\
BR$(\tau\to e\gamma)$ & $2.0\times 10^{-22}$ & $2.5\times 10^{-22}$ & $2.4\times 10^{-22}$ \\
BR$(\tau\to \mu\gamma)$ & $3.4\times 10^{-18}$ & $4.2\times 10^{-18}$ & $4.0\times 10^{-18}$ \\
BR$(\mu\to 3e)$ & $7.8\times 10^{-26}$ & $9.6\times 10^{-26}$ & $1.5\times 10^{-25}$ \\
BR$(\tau\to 3e)$ & $2.3\times 10^{-24}$ & $2.9\times 10^{-24}$ & $3.1\times 10^{-24}$ \\
BR$(\tau\to 3\mu)$ & $1.3\times 10^{-20}$ & $1.6\times 10^{-20}$ & $1.6\times 10^{-19}$ \\ \hline
$|\eta_{ee}|$ & $5.26\times 10^{-5}$ & $4.02\times 10^{-5}$ & $3.92\times 10^{-4}$ \\
$|\eta_{\mu\mu}|$ & $1.50\times 10^{-6}$ & $1.50\times 10^{-6}$ & $1.50\times 10^{-6}$ \\
$|\eta_{\tau\tau}|$ & $1.50\times 10^{-6}$ & $1.50\times 10^{-6}$ & $1.50\times 10^{-6}$ \\ \hline\hline
\end{tabular}
\caption{All relevant model input parameters, Higgs mass, DM mass, 
relic density, spin-independent cross section and other relevant 
low-energy flavor sector observables for the three chosen BPs.}
\label{tab:low}
\end{table}
We fit the neutrino masses and mixing angles with an off-diagonal $\mu_s$. While fitting the neutrino oscillation data, 
we assume normal hierarchy in neutrino mass for the present study. However, the inverted hierarchy scenario can also 
be fitted equally with different entries in the $\mu_s$ matrix. These entries being of keV order at most, do not 
affect sneutrino masses and mixing angles in any significant manner. Hence even with an inverted hierarchical mass structure of 
the neutrinos, the DM analysis remains unchanged. 
Following are the values for $\mu_S$ for the three BPs shown above:
\begin{eqnarray}
\mu_S = \left\{\begin{array}{cc}
\left(\begin{array}{ccc}
38.51 & 463.88 & 150.49 \\
463.88 & 8931.54 & 7309.49 \\
150.49 & 7309.49 & 9661.64
\end{array}\right)~{\rm eV} &~~~~~ ({\rm BP1}) \\
\left(\begin{array}{ccc}
50.46 & 530.34 & 172.04 \\
530.34 & 8907.71 & 7289.99 \\
172.04 & 7289.99 & 9635.86
\end{array}\right)~{\rm eV} &~~~~~ ({\rm BP2}) \\
\left(\begin{array}{ccc}
5.16 & 169.68 & 55.04 \\
169.68 & 8909.98 & 7291.84 \\
55.04 & 7291.84 & 9638.31
\end{array}\right)~{\rm eV} &~~~~~ ({\rm BP3}) 
\end{array}\right. 
\end{eqnarray}
All the annihilation cross-sections corresponding to the dominant decay modes are provided in Table \ref{tab:sigv_contbn} for the three benchmark points.
For all the points, annihilation into $b\bar b$ final state is the dominant one, followed by the $\tau\bar\tau$ and $c \bar c$ channels
which are down by one order of magnitude, relative to $b\bar b$ channel. For all the benchmark points, $<\sigma v>_{b\bar b}$ values are roughly half 
of the limit put by \cite{Daylan:2014rsa}. For all these points we can fit the galactic center gamma ray 
excess perfectly well (Fig.~\ref{spec_fit}).   
\begin{table}[h!]
\begin{center}
\begin{tabular}{||c|c c c||}\hline\hline
                &  & $<\sigma v>_{b\bar b}~({\rm cm^3 s^{-1}})$  & \\
contributing channels & BP1 & BP2 & BP3 \\ 
\hline\hline
$\widetilde\nu_1 \widetilde\nu_1 \to b\bar b$ & $4.89\times 10^{-27}$ & $4.93\times 10^{-27}$ & $5.00\times 10^{-27}$ \\ 
$\widetilde\nu_1 \widetilde\nu_1 \to \tau\bar\tau$ & $8.59\times 10^{-28}$ & $8.66\times 10^{-28}$ & $8.78\times 10^{-28}$ \\
$\widetilde\nu_1 \widetilde\nu_1 \to c\bar c$  & $< 10^{-28}$ & $< 10^{-28}$ & $< 10^{-28}$ \\
\hline\hline 
\end{tabular}
\end{center}
\caption{The annihilation cross-sections for different annihilation channels for the three BPs.}
\label{tab:sigv_contbn}
\end{table}

\subsection{Anti-proton Flux:}
\begin{figure}[h!]
\centering
\includegraphics[trim = 0 100 10 100, clip,height=10cm,width=9cm]{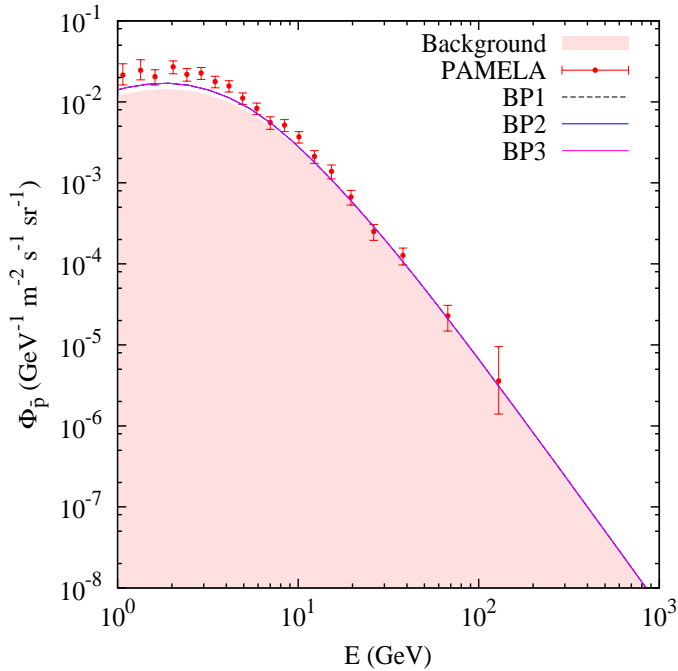}
\caption{The anti-proton flux plotted as a function of energy. 
The pink(gray) patch is the background calculated using GALPROP. The dashed and solid lines indicate the signal contributions of BP1, BP2 and BP3 summed together 
bin wise with the background respectively. The points with the error bars 
correspond to the PAMELA data.}
\label{fig:antip_flux}
\end{figure}

The satellite-based experiment PAMELA\cite{Adriani:2008zq} has measured the cosmic ray antiproton
flux and has seen that it is consistent with the secondary production of anti-protons due to
cosmic ray propagation.
As mentioned earlier, an enhancement in $b\bar b$ annihilation cross-section of the sneutrino DM
can also lead to an increase in anti-proton flux.
To test our benchmark points in the light of anti-proton flux in the final state, we generate the cosmic-ray
background contribution using GALPROPv54 \cite{Strong:1998pw} and combine them with our signal contribution. The input parameters for GALPROP that we choose are as follows \cite{Ackermann:2010ij}: Electron injection index of 2.5 for E $>$ 4 GeV and 1.6 for E $\leq$ 4 GeV with a modulation potential
of 550 MV, a spatial Kolmogorov diffusion with diffusion
coefficient of $5.75\times10^{28}~{\rm cm^2. s^{−1}}$ and spectral index of
0.33 and we also choose Alfven
speed of 30 $\rm km~ s^{-1}$, and halo radius of 8.5 kpc.\\
As can be seen from Fig.~\ref{fig:antip_flux} that the experimental data provided by the PAMELA
shows very little excess over the generated background. But the sneutrino annihilation contribution combined
with the background fits the experimental data pretty well. We show the distribution for all the three benchmark points.

Fitting the galactic center gamma-ray excess really constrains the sneutrino sector parameters, specially, 
$M_R$, $y_{\nu}$, $B_{M_R}$ and $A_{\nu}$. Among these parameters, $M_R$ and $y_{\nu}$ also affect the 
neutrino mass in the model. Hence using this gamma-ray excess fit, one can constrain the choice of parameters 
in the neutrino sector also. Of course, as the benchmark points suggest, there can be various choices of $M_R$ and $y_{\nu}$ 
that can simultaneously fit the excess and other DM and collider experimental bounds besides fitting the neutrino oscillation data. 
Before we conclude, we would like to make few comments about the possibilities 
of studying this scenario at the LHC . 
In this scenario, one expect to observe various missing energy signals arising 
from the LSP sneutrinos to determine most suitable parameters space of this
model. Some collider related studies of this model 
have been already performed 
\cite{BhupalDev:2012ru,Banerjee:2013fga,Mondal:2012jv}. 
Our results strongly suggests that this fit can be used as a constraint 
on the model parameters which can be further tested at collider experiments.

\section{Conclusions:}\label{concl}
The Fermi-LAT has observed an excess in gamma-ray flux in the galactic center with latitude $|b|<5^{\circ}$
that can be explained by secondary photons originating from DM annihilation into charged fermionic final states.
In this paper we try to fit this observed gamma-ray excess in sneutrino DM scenario of the supersymmetric
inverse seesaw model(SISM), where, a pair of the sneutrino DM dominantly annihilate 
via s-channel Higgs mediated process to $b \bar b$ states which eventually leads to the secondary gamma-rays in 
the final state. We perform a detailed $\chi^2$ analysis with the relevant model parameters
to fit the observed gamma ray spectrum. Our findings show that the excess in the
photon energy bump as seen by the Fermi-LAT collaboration can indeed be explained in SISM with a sneutrino DM of mass 
close to 40 GeV and a good fit to the spectrum requires $<\sigma v>_{b \bar b}$ 
$\sim 5\times 10^{-27}~{\rm cm^3.s^{-1}}$. However, one should note that similar particle physics studies  
\cite{Hooper:2013rwa,Daylan:2014rsa} to explain this phenomena indicate towards a $<\sigma v>_{b\bar b}$ 
which is roughly twice the rate that we obtain in our analysis. 

It is interesting to note that the SISM predicts the existence of very light 
Higgs boson of mass 86.7 GeV which is allowed by all the present experimental 
limits and it may even escape detection 
at the 14 TeV LHC as shown in a similar light Higgs scenario study \cite{Bhattacherjee:2013vga}. However, at the future $e^+e^-$ international 
linear collider (ILC) it may be possible to pin down this ultra light Higgs scenario and
the detailed signal and background analysis in this context will be reported elsewhere \cite{dkg:sm:is}.

We show the spectrum fit for three different benchmark points over the whole parameter space. While 
doing this analysis, we also check the anti-proton flux in those 
particular benchmark points and we find it to be consistent with the 
PAMELA experimental data. 
The obtained annihilation cross-section in this study may be used in addition to the other DM constraints 
to reduce the model parameter space to a great deal. 
Moreover, since these parameters are instrumental for both the neutrino sector as well as the DM sector studies, 
collider search into this parameter space can provide indirect probe for both the experimental findings if SISM is indeed the mechanism 
for neutrino mass generation. 
\section{Acknowledgment}
SM wishes to thank the Department of Science and Technology, Government of 
India for a Senior Research Fellowship. Authors would like to thank S. Roy,
P.S.B. Dev, N. Okada, G.~Belanger and Arindam Chaterjee for helpful discussions. 
DKG \& IS would like to thank RECAPP, HRI for the hospitality where part of 
this work was done.

\end{document}